\documentclass{llncs}

\usepackage{xspace,amsmath,url}
\usepackage{color}
\usepackage{booktabs}
\usepackage[utf8]{inputenc}
\usepackage{comment}
\usepackage{cite}
\usepackage{url}
\usepackage{graphicx}

\newcommand{\exclude}[1]{}

\def\mmod{~\textrm{mod}~}

\newcounter{lnoc}
\newenvironment{sdalgorithm}[1]{%
\hrule height 0.8pt \vspace{0.6ex} \small#1\vspace{0.6ex}\hrule height 0.5pt \vspace{-2.0ex}
\setcounter{lnoc}{0}
\small
\begin{tabbing}
000000\=XXI\=XXI\=XXI\=XXI\=XXI\=XXI\=\kill
}{%
\end{tabbing}
\vspace{-2.0ex}\hrule height 0.8pt\vspace{0ex}}
\newcommand{\lno}[1][0]{{\footnotesize\sffamily 
\ifnum#1=0
\stepcounter{lnoc} 
\ifnum\thelnoc<10
\phantom0%
\fi
\thelnoc
\else
\thelnoc.#1
\fi
}\>}

\newcommand{\pcdo}{{\bfseries do~}}
\newcommand{\pcif}{{\bfseries if~}}
\newcommand{\pcthen}{{\bfseries then~}}
\newcommand{\pcelse}{{\bfseries else~}}

\newcommand{\pcwhile}{{\bfseries while~}}

\newcommand{\pcand}{{\bfseries and~}}

\newcommand{\pcreturn}{{\bfseries return~}}

\usepackage{hyperref}

\begin{document}

\title{FM-index for dummies}

\author{Szymon Grabowski$^\dag$, Marcin Raniszewski$^\dag$ 
        and Sebastian Deorowicz$^\ddag$}
\institute{
$^\dag$ Lodz University of Technology, Institute of Applied Computer Science,\\
  Al.\ Politechniki 11, 90--924 {\L}\'od\'z, Poland, \email{(sgrabow|mranisz)@kis.p.lodz.pl}\\
$^\ddag$ Institute of Informatics, Silesian University of Technology,\\ 
  Akademicka 16, 44-100 Gliwice, Poland, \email{sebastian.deorowicz@polsl.pl}\\
}

\maketitle

\begin{abstract}
The FM-index is a celebrated compressed data structure for full-text pattern 
searching.
After the first wave of interest in its theoretical developments, 
we can observe a surge of interest in practical FM-index variants in the last 
few years.
These enhancements are often related to a bit-vector representation, 
augmented with an efficient rank-handling data structure. 
In this work, we propose a new, cache-friendly, implementation of
the rank primitive and advocate for a very simple architecture of the FM-index,
which trades compression ratio for speed.
Experimental results show that our variants are 2--3 times faster 
than the fastest known ones, for the price of using typically 1.5--5 times more space.
\end{abstract}

\section{Introduction}
The rapid development of compressed data structures 
in the first decade of our century changed the landscape 
of modern algorithmics.
Prominent examples of those achievements are compressed 
indexes for unstructured~\cite{GV00,FM00,MN07} and semi-structured texts~\cite{FLMM09}, 
compressed trees~\cite{BDMRRR05}, graphs~\cite{BV04,BLN14}, 
binary relations~\cite{BHMR11}, RDF triples~\cite{BCFN15} 
and color range counting~\cite{N14}.
Real applications of these sophisticated data structures 
however lag behind, with a notable exception of bioinformatics~\cite{VDFD12,DG13}.
In this work we revisit one of the most celebrated concepts 
in stringology in recent years, the FM-index by Ferragina and Manzini~\cite{FM00,FM05}.
The key component of virtually any of multiple variants 
of this index is the operation {\em rank}, usually performed on a 
bit-vector $B$, which, for a given integer $j$, returns the number 
of set bits in $B$'s prefix of length $j$.
We propose a new, cache-friendly, implementation of the rank primitive 
and advocate for a very simple architecture of the FM-index, 
which trades compression ratio for speed.

The following notation will be used throughout the paper.
An index will be built for a text $T[1 \ldots n]$ 
over an integer alphabet $\Sigma = \{1, \ldots, \sigma\}$.
The index will be queried with patterns of the form $P[1 \dots m]$.
The rank operation will be calculated for the bit vector 
$B[1 \ldots n]$.
We assume the CPU cache line is of size $L = 512$ (bits).
(As the experiments are run on an Intel Core CPU, some variants are optimized 
in terms of {\em pairs} of successive cache lines, i.e., blocks of 1024 bits,
which allows to use Streamer, a second-level cache prefetcher; 
more details in~\cite[Sect.~3.7.3]{IntelManual}.)
All logarithms are in base 2.
The colloquial term ``popcount'' (population count) will often be used 
for the operation of counting the number of bits 1 in a given bit sequence.

\section{The FM-index architecture} \label{sec:arch}

The FM-index is basically the result of the 
Burrows--Wheeler transform (BWT) of text $T$, denoted as $T^\mathrm{bwt}$,
with two helper structures:
a count array 
$C[1 \ldots \sigma]$ 
such that 
$C[i] = |\{T[j]: T[j] \leq i~\text{and}~1 \leq j \leq n\}|$, 
and a data structure answering 
$\mathit{Occ}(T^\mathrm{bwt},c,pos) = |\{T^\mathrm{bwt}[j]: T^\mathrm{bwt}[j] = c~\text{and}~1 \leq j \leq pos\}|$ queries.
This allows to count the occurrences of a pattern $P$, finding the ranges 
of the (implicit) suffix array (SA) for $T$ starting with successive suffixes 
of $P$ in the 
successive 
loop iterations, see Fig.~\ref{alg:FMsearch}.
If the function $Occ(\cdot)$ is realized with a wavelet tree~\cite{MN07,Nav13}, 
the count queries run in $O(m\log\sigma)$ worst case time.
Plenty of FM-index variants exist, with different space-time complexity tradeoffs 
and different practical performances.
For a survey, see~\cite{MN07}; important recent results were presented 
in~\cite{KP11,BN14}.
Experimental comparisons can be found, e.g., in~\cite{FGNV09,GP14}.

\begin{figure}[t]
\begin{sdalgorithm}{Count-Occs($T^\mathrm{bwt}$, $n$, $P$, $m$)}
\lno	$i \gets m$\\
\lno	$\mathit{sp} \gets 1$; $\mathit{ep} \gets n$ \\
\lno	\pcwhile $(\mathit{sp} \leq \mathit{ep})$~\pcand $(i\geq 1)$ \pcdo \\
\lno	\> $c \gets P[i]$ \\
\lno	\> $sp \gets C[c]+\mathit{Occ}(T^\mathrm{bwt},c,\mathit{sp}-1)+1$ \\
\lno	\> $ep \gets C[c]+\mathit{Occ}(T^\mathrm{bwt},c,\mathit{ep})$ \\
\lno	\> $i \gets i-1$ \\
\lno	\pcif $(\mathit{ep} < \mathit{sp})$ \pcthen \pcreturn ``not found'' \pcelse \pcreturn ``found ($\mathit{ep}-\mathit{sp}+1$) occs''
\end{sdalgorithm}
\caption{Counting the number of occurrences of pattern $P$ in $T$ with the FM-index.}
\label{alg:FMsearch}
\end{figure}

\section{Rank with one cache miss} \label{sec:rank1}
Jacobson~\cite{J89} showed that the rank operation for a bit vector 
of length $n$ can be implemented in constant time 
using $O(n\log\log n/\log^2 n)$ extra bits.
This however requires three memory accesses (to one superblock counter 
and one block counter, plus a lookup into a table 
with precomputed popcount answers), therefore more practical 
ideas were later presented~\cite{GGMN05,V08,GP14}.
Following the idea of Gog and Petri~\cite{GP14} 
(who in turn extended the approach of Vigna~\cite{V08}),
we use one level of counters, interleaving the bit vector data 
with the counters, 
to improve the locality of memory accesses.
If one counter and an interval of bits from $B$ takes exactly 
one (aligned) cache line, we can calculate the rank with 
one cache miss in the worst case.
In contrast, Gog and Petri~\cite{GP14} interleave 64-bit counters with 
bit vector data of 256 bits in their \textsc{Rank-1L} variant. 
Note that their structure is logically divided into chunks of 
$64 + 256 = 320$ bits, which are usually not aligned to the cache line.

We come back to our variant.
Assume for clarity that $n$ is a multiple of $512 - 64 = 448$.
More precisely, we maintain a bit table $B'[1 \ldots n']$, 
where $n' = 512n/448$,
and $B'[512 i + 1 \ldots 512 i + 64]$ is a 64-bit counter $R[i]$ 
storing the value of $rank_1(B, 448 i)$,
while $B'[512 i + 65 \ldots 512 i + 512] = B[448 i + 1 \ldots 448 i + 448]$, 
for any valid $i \geq 0$.
Now, 
$rank_1(B, j) = R[\lfloor j/448 \rfloor] + popcnt(B'[512 \lfloor j/448 \rfloor + 65 \ldots 512 \lfloor j/448 \rfloor + 65 + (j \mmod 448)])$.
The popcount operation is performed using the hardware 64-bit opcode 
POPCNT (known as the \texttt{\_\_builtin\_popcountll} function in gcc), 
which seems fastest on the Intel Nehalem and later CPUs.

Note that for $n < 2^{32}$ 32-bit counters are enough, 
yet using 64-bit counters provides proper alignment for 
calling the \texttt{\_\_builtin\_popcountll} instruction (7 times).
Alternatively, we could use a 32-bit counter, then call 
the 32-bit \texttt{\_\_builtin\_popcount} once and finally 
\texttt{\_\_builtin\_popcountll} 7 times.
Yet another pair of variants (with 64-bit and 32-bit counters, 
respectively), reducing the number of popcount instructions, 
but using more space, maintains 256-bit rather than 512-bit blocks in $B'$.
The space overhead of the four possible variants 
is $64 / 448 = 14.3\%$, $32 / 480 = 6.7\%$, 
$64 / 192 = 33.3\%$ and $32 / 224 = 14.3\%$, respectively.

A drawback of our approach is that a division by a number not being 
a power of 2 (e.g., 448) is required in a rank computation.
On the other hand, modern compilers (including gcc) convert 
integer division by a constant into a multiplication and a few additions 
and shifts~\cite[Chap.~10]{HackersDelight}, 
which is several times faster than general division.

We also note that our bit table $B'$ must be aligned to a multiple of 
the cache line size (failing to do so does not guarantee a single cache 
miss).

Additionally, we employ software prefetching to reduce the access time 
to a memory cell.
In the main loop (lines 3--7 in Fig.~\ref{alg:FMsearch}), 
for each pattern symbol it is necessary to determine the 
new left and right boundaries of the current range of the (implicit) suffix array.
Just when we obtain the new left boundary we make use of software prefetching 
to bring the necessary address to the cache.
The same is done for the right boundary.
The amount of calculations between the prefetch and the access 
to the cell is rather small 
(determination of the opposite boundary and a few additions and array accesses), 
so the gain in time is rather moderate, yet noticeable.

\section{FM-dummy} \label{sec:dummy}

Dealing with the dependence on the alphabet size is one of the 
key issues in FM-index design.
We propose several variants of the FM-index.
Although these ideas are hardly novel, surprisingly we are not 
aware of their implementations. 
In the experimental section we will however show that these schemes, 
together with our rank implementations, offer attractive time-space tradeoffs.

In the first variant, for a small alphabet, 
we maintain $\sigma$ bit vectors of length $n$, one per alphabet symbol, 
together with the corresponding rank data.
We propose to use it if $\sigma \leq 16$.
Let us denote this algorithm as \texttt{FM-dummy1}.
We admit that this scheme, with compressed rank, 
was proposed by M{\"a}kinen and Navarro 
in 2004 in a technical report~\cite{MN04tr},
in Section 3.2 appropriately entitled ``Replacing {\em Occ} Structure 
by Individual Bit Arrays''.
Their idea was to obtain $O(m)$ count time (i.e., with no dependence 
on the alphabet size) with $O(H_0 n)$ bits of space, 
yet in an erratum note dated 9th Dec. 2004 they noticed an error in analysis.
In theoretical terms, the desired properties of this algorithm are obtained 
only for $\sigma = O(\operatorname{polylog}(n))$.
The same solution is also used in ABySS~\cite{simpson2009abyss}, 
a well-known de novo genome assembler.\footnote{As pointed out 
to us by Shaun D.\ Jackman, one of ABySS's authors (June 2015).}

The second variant is suggested for the case of $\sigma > 16$.
Before applying the BWT, we encode the text using a dense code.
First we use the $(s,c,b,o)$-DC (SCBDC)~\cite{FNP11} scheme, 
which is both prefix- and suffix-free and thus requires no verifications.
We use this code on nybbles, i.e., set the parameters in a way to have 
$s + c + b + o = 16$.
This means that the code space of 16 values is divided into disjoint subsets 
of sizes:
$(i)$, $o$, the number of length-1 codewords, 
$(ii)$, $b$, the number of distinct prefixes of codewords (``beginners'') of length $\geq 2$,
$(iii)$, $s$, the number of distinct suffixes of codewords (``stoppers'') of length $\geq 2$,
and 
$(iv)$, $c$, the number of distinct middle symbols of codewords (``continuers'') of length $\geq 3$.
The number of codewords of length up to $j \geq 3$ is 
$o + \sum\limits_{i=0}^{j-2} b s c^i$.
For example, setting $(s,c,b,o) = (4,2,4,6)$ for \texttt{english.200MB} 
we obtain the average codeword length of 1.584 (nybble per symbol).
We denote this variant as \texttt{FM-dummy2}.

The other option is to use a simpler (and denser) encoding, 
with beginners and continuers only,
of counts $b$ and $c$, respectively, where $b + c = 16$ 
or $b + c = 8$ (these versions are denoted in the experiments 
with `\_4' or `\_3', respectively, in their names),
which produces $\sum\limits_{i=0}^{j-1} b c^i$ codewords of length 
up to $j \geq 2$ (cf.~\cite[Sect.~4]{G08}).
There are two issues with this encoding though:
$(i)$ any match in the encoded text (except at the very end of the text!) 
must be followed with a symbol from the beginners, 
hence the (backward) search over the pattern must start with a dummy 
``any-beginner symbol'' in the FM-index; 
fortunately it is easy to simulate it with setting 
appropriately the initial suffix range (the {\em sp} and {\em ep} 
variables in Fig.~\ref{alg:FMsearch}),
$(ii)$ {\em bidirectional} searches over the FM-index, 
which have applications in approximate index string matching~\cite{LLTWWY09}
and some DNA sequence analysis problems (e.g., maximal unique matches)~\cite{BCKM13},
become problematic.
In the experimental section this modification is dubbed \texttt{FM-dummy2cb}.

In \texttt{FM-dummy1} and both versions of \texttt{FM-dummy2} we also 
try out eliminating part of the linear scan (several popcounts) over a block.
More precisely, with blocks of 256 bits we use 48 bits for the counter 
and two bytes of its 64-bit word are spent for storing the ranks 
for the 64- and the 128-bit prefix of the block data (which reduces the 
maximal number of POPCNT operations in a block from 3 to 1).
Similarly, with blocks of 512 bits we use 40 bits for the counter 
and three bytes storing the number of ones in three successive subblocks 
of 128 bits each.
Implementations involving this idea have letter `c' in the name, e.g., 
\texttt{FM-dummy1\_256c}.

DNA is an important application of text indexes 
and the \texttt{FM-dummy1} variant presented above may not always be preferred 
since it is not quite succinct.
To address this issue, we propose \texttt{FM-dummy3}, which assumes 
the alphabet of size 5 (\texttt{ACGTN}), where the symbol \texttt{N} stands 
for any symbol not from \texttt{ACGT} in the text.
It is also assumed that the patterns are from the \texttt{ACGT} (sub)alphabet, 
otherwise they would make little sense from the biological point of view.
In a block of 512 (1024) bits, there are four 32-bit counters for the 
four valid pattern symbols, 
followed by 384 (896) bits of data.
The block data consist of symbols packed into bytes in triples 
(which can be easily done, since $5^3 \leq 256$).
To obtain a rank for a given symbol and a given position in block, 
we scan the data bytes with a reference to a lookup table 
having $125 \times 4$ entries.

Finally, we implemented an FM-index with a Huffman-shaped multiary wavelet tree, 
namely with arity 4 and 8 (\texttt{FM-HWT4} and \texttt{FM-HWT8}).
(For completeness, we added also a variant with a Huffman-shaped binary 
wavelet tree.)
In the 4-ary (8-ary) case, 
each block contains 4 (8) 32-bit counters, followed by  
packed data as a sequence of pairs (triples) of bits.
The block size is a parameter, set to 512 or 1024 bits.
For example, if the 8-ary variant is chosen and 512-bit blocks, 
we have $512 - 8 \times 32 = 256$ bits for the data, 
which are grouped in four 64-bit words, each containing 21 triples of bits 
(1 bit per 64-bit word is then ``wasted'').
Counting the rank for a symbol from the 8-ary alphabet for the data sequence 
is performed using simple bitwise operations, including 
\texttt{xor}, \texttt{and} and shifts, 
followed by the hardware popcount.

\section{Boosting short pattern search with a hash table} \label{sec:ht}

In~\cite{GrabowskiR14} we showed how to augment the standard suffix array with 
a hash table (HT), to start the binary search from a much more narrow interval. 
The start and end position in the suffix array for each range of suffixes 
having a common prefix of length $k$ was inserted into the HT,
with the hash function calculated for the prefix string.
The same function was applied to the pattern's prefix and after a HT lookup 
the binary search was continued with reduced number of steps.
The mechanism requires $m \geq k$.

Later, we incorported this idea in another full-text index, SamSAMi~\cite{GrabowskiR15}, 
and now propose to use it with an FM index.
First, the pattern's {\em suffix} of length $k$ is sought in the HT 
and then the search continues in a standard manner.
The number of symbols submitted to a standard FM-index backward search 
is reduced from $m$ to $m - k$.
Each entry of the hash table stores the corresponding $k$ symbols 
and two integers, for the left and the right boundary of the suffix array interval.
Note that, contrary to the SA-hash solution~\cite{GrabowskiR14}, we cannot 
avoid storing the $k$ symbols since we don't have an explicit 
suffix array and fast access to arbitrary text position, to resolve collisions.
Note also that using a perfect hashing scheme does not fix this issue, 
since looking for a $k$-gram {\em not occurring in the text} gives a ``random'' 
position in the hash table, which could imply spurious matches.

The size of the hash table component for a fixed $k$
depends of course on the dataset used, but also if we use the whole 
alphabet 
for the DNA dataset, or assume that the patterns
are over the \texttt{ACGT} subalphabet 
(\texttt{FM-dummy1} and \texttt{FM-dummy3} variants).
In practice, however, this effect is negligible.

In the experimental section we set $k = 5$ and run experiments only on short 
patterns (of length from 6 to 10).
This allows to speed up searches significantly for a price of only moderate 
increase in the space use for real texts.

\section{Experimental results}
All experiments were run on a machine equipped with 
a 6-core Intel i7 CPU 
(4930K) 
clocked at 3.4\,GHz, with 64\,GB of RAM, 
running Ubuntu 14.04 LTS 64-bit.
The RAM modules were $8 \times 8$\,GB DDR3-1600 with the timings 
11-11-11 (Kingston KVR16R11D4K4/64).
The CPU cache sizes were: 
$6 \times 32$\,KB (data) and $6 \times 32$\,KB (instructions) in the L1 level, 
$6 \times 256$\,KB in L2 
and 12\,MB in L3.
One CPU core was used for the computations.
All codes were written in C++ and compiled with 64-bit gcc 4.8.2, 
with \texttt{-O3} option
(and for the search algorithms with the additional \texttt{-mpopcnt} option).
The source codes for our implementations 
are available at
\url{https://github.com/mranisz/fmdummy/releases/tag/v1.0.0}.

\begin{figure}
\centerline{
\includegraphics[width=0.49\textwidth,scale=1.0]{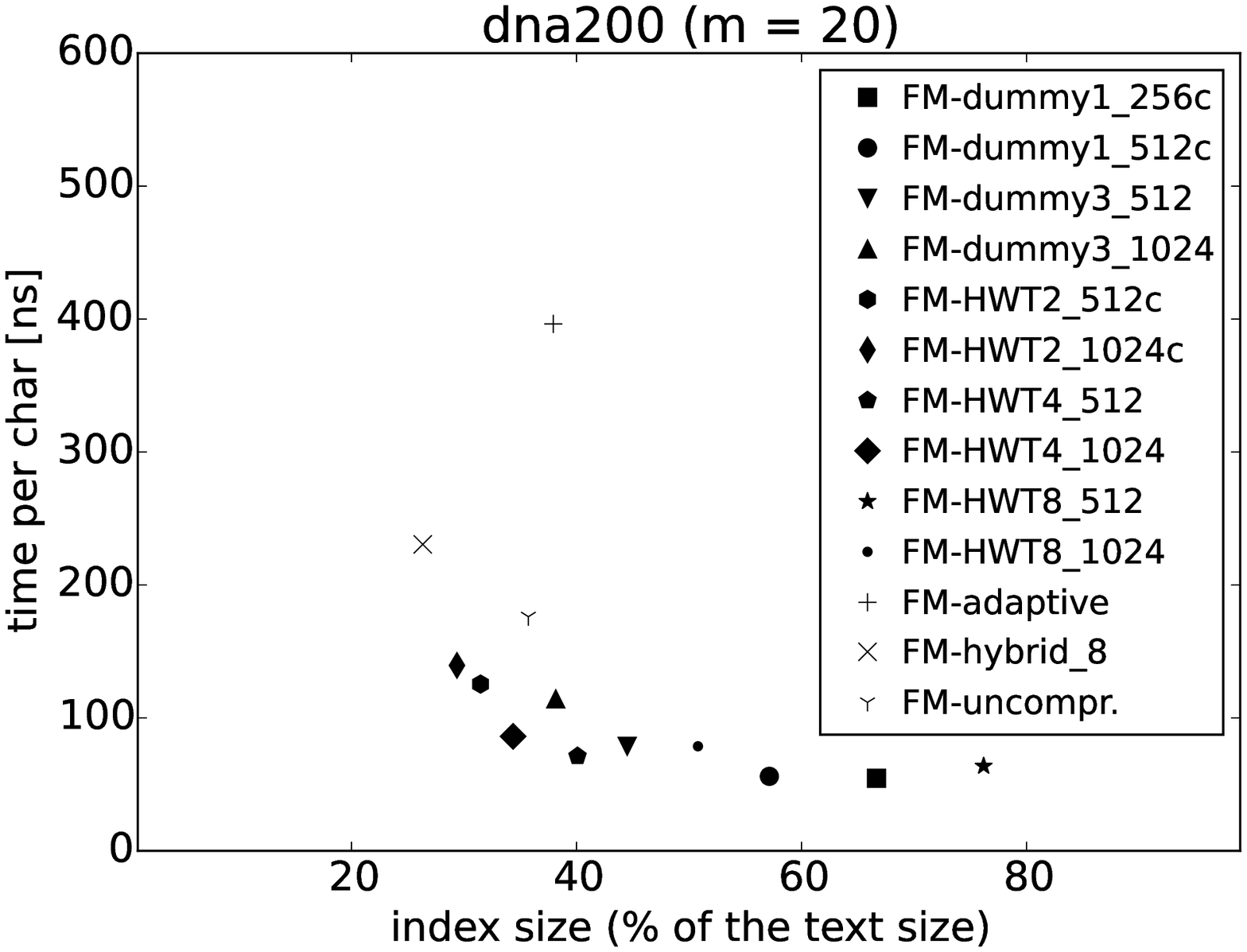}
\includegraphics[width=0.49\textwidth,scale=1.0]{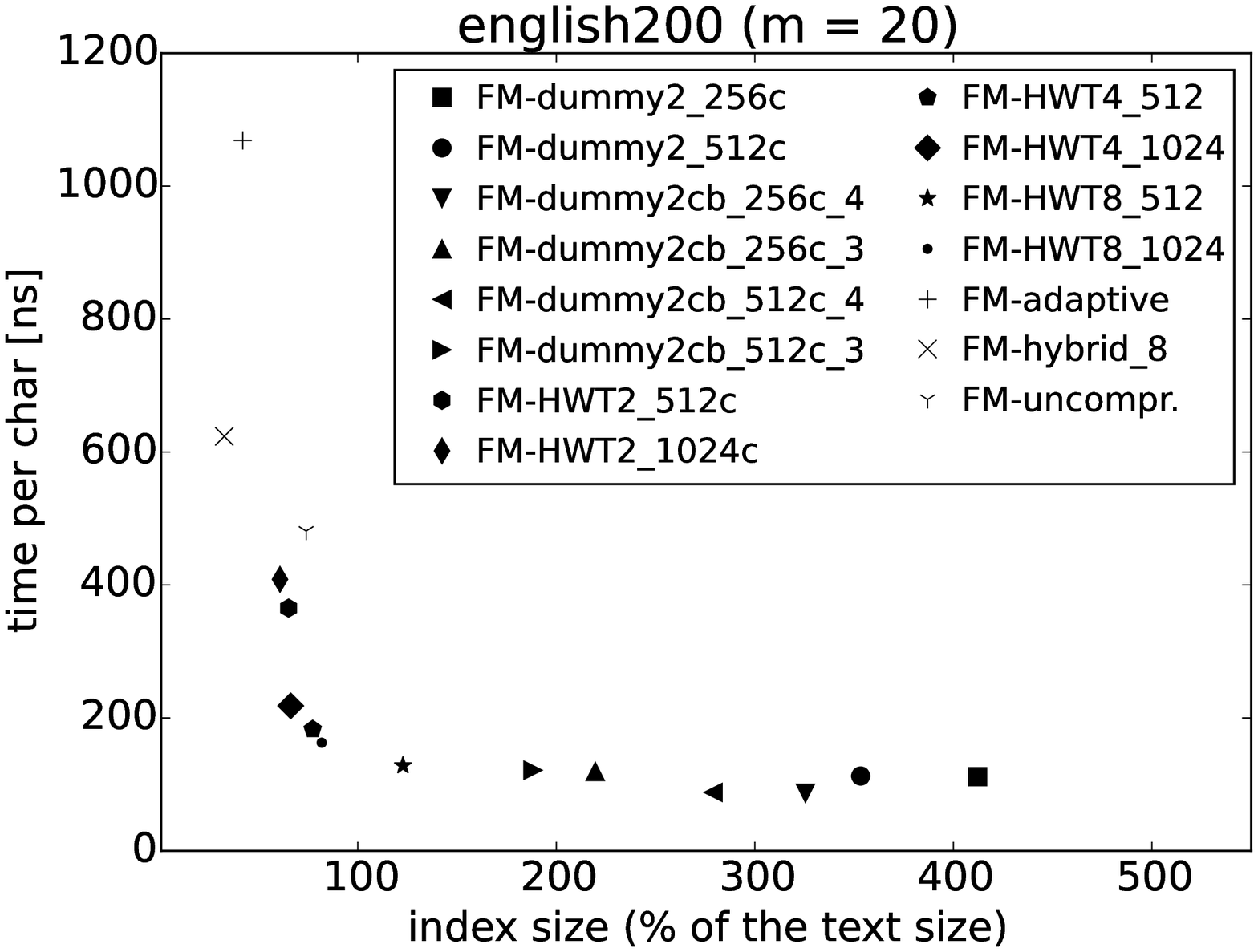}
}
\centerline{
\includegraphics[width=0.49\textwidth,scale=1.0]{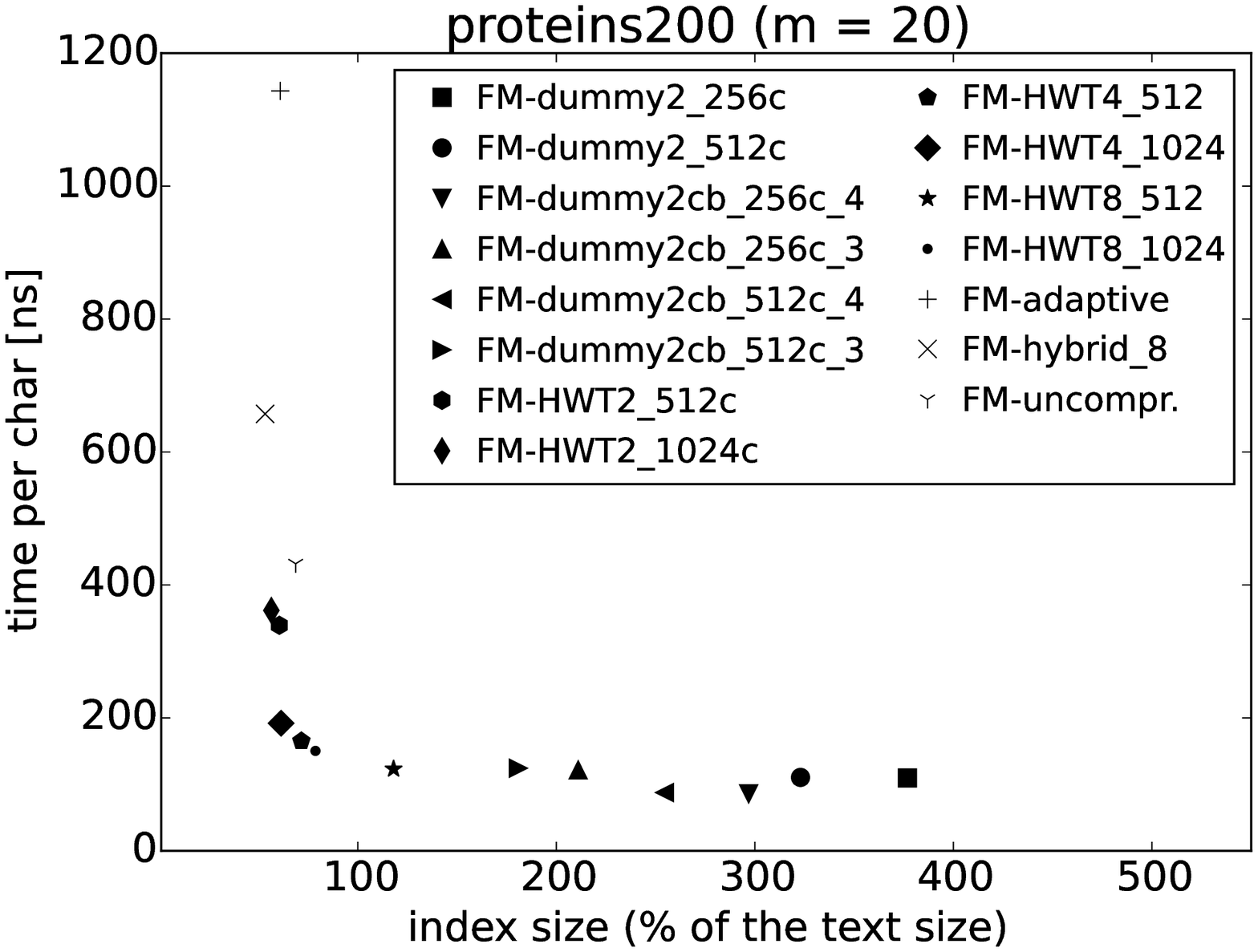}
\includegraphics[width=0.49\textwidth,scale=1.0]{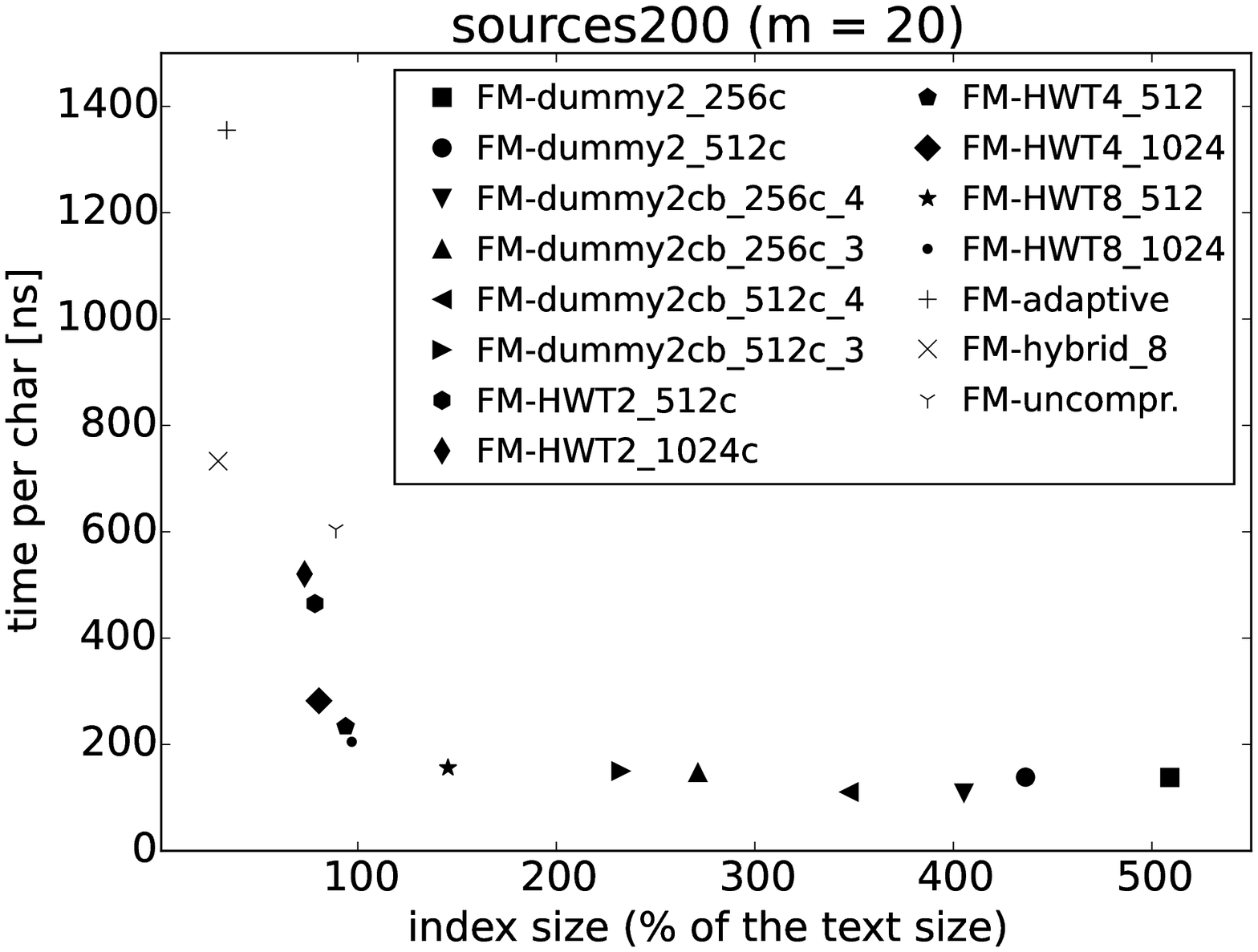}
}
\centerline{
\includegraphics[width=0.49\textwidth,scale=1.0]{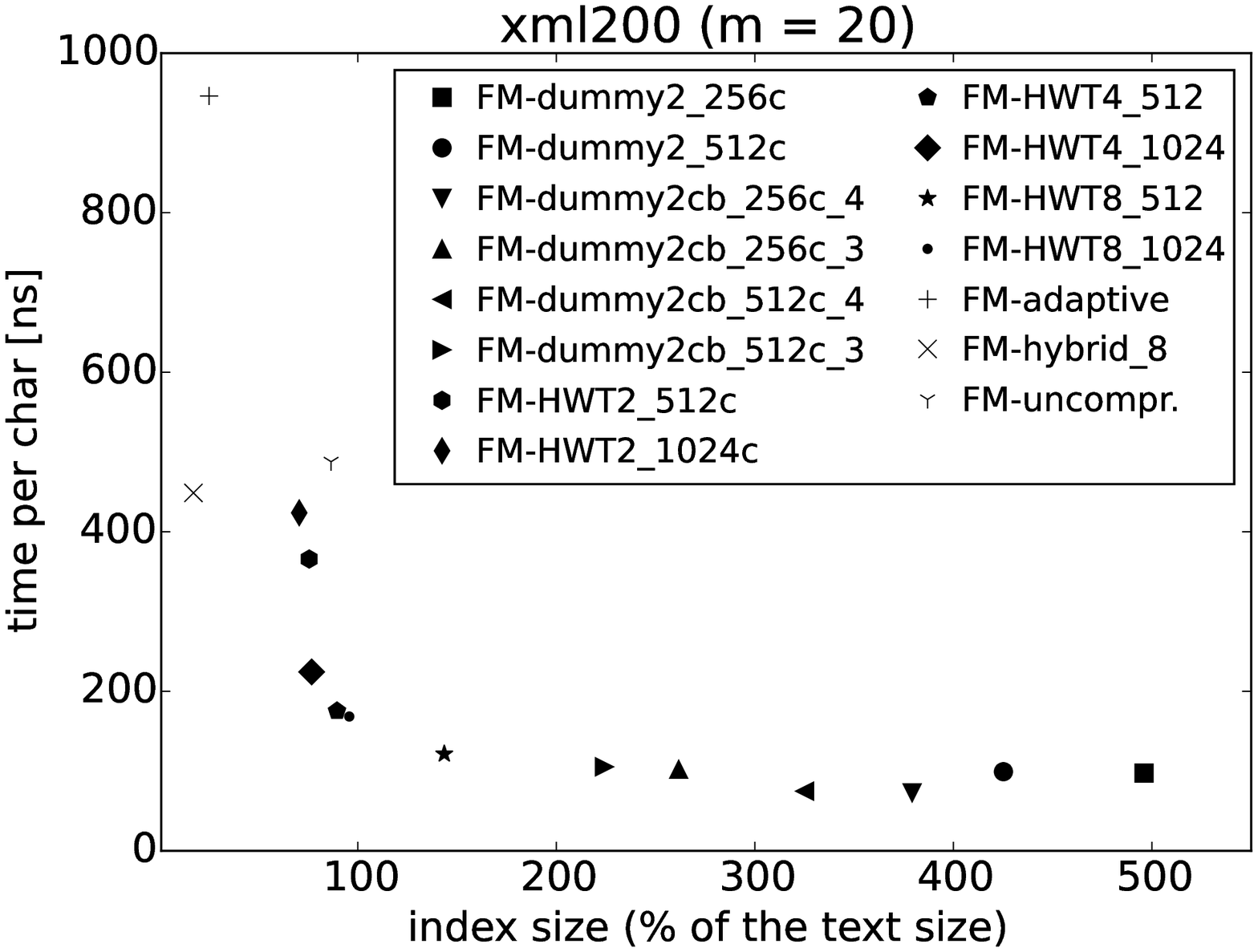}
}
\caption[Results]
{Count query. 1M patterns of length 20 were used. 
Times are averages in ns per character.
The patterns were extracted from the respective texts.}
\label{fig:times}
\end{figure}

The test datasets were taken from the Pizza~\&~Chili 
site\footnote{\url{http://pizzachili.dcc.uchile.cl/}}.
We used the 200-megabyte versions of the files \texttt{dna}, 
\texttt{english}, \texttt{proteins}, \texttt{sources} and \texttt{xml}.

In order to evaluate the search algorithms, we generated 1 million patterns 
of length $m = 20$;
the patterns were extracted randomly from the 
corresponding datasets (i.e., each pattern returns at least one match), 
with a special procedure for the DNA dataset, where only patterns 
over the subalphabet \texttt{ACGT} were allowed.
The performance of count queries only was measured.
Actually, the current implementations of our indexes have no support 
for the locate query, but for a possibly honest comparison we reduced 
the sampling in the other indexes (where it is available) 
to very large values (at least 1 million), in order to make the space overhead 
totally negligible.

We compared the following FM-index variants:
\begin{itemize}
\item \texttt{FM-uncompressed}\footnote{\url{https://github.com/simongog/sdsl-lite}}, 
based on a Huffman-shaped binary wavelet tree 
with uncompressed bit-vectors; it is called V5 in~\cite{GP14},

\item \texttt{FM-hybrid}\footnote{\url{https://www.cs.helsinki.fi/group/pads/hybrid_bitvector.html}},
being the same as above but with the wavelet tree bit-vectors 
divided into blocks for which one of three simple encoding methods is 
separately chosen~\cite{KKP14}; the superblock size of 8 was always chosen, 
as the sizes of 16, 32 and 64 gave similar results (the index with parameter 8 
was the fastest yet using slightly more space than the other choices),

\item \texttt{FM-adaptive}\footnote{\url{https://github.com/chenlonggang/Adaptive-FM-index}},
a recently proposed algorithm~\cite{HCZVNY15} 
related to~\cite{KKP14}, 
yet with the main modification of using a variable-length coding (Gamma coding) 
in blocks rather than fixed length coding,

\item \texttt{FM-dummy1}, our variant for a small alphabet (DNA only from the test collection), 
with block sizes of 256 or 512 bits,

\item \texttt{FM-dummy3}, our variant for DNA specifically, with block sizes of 
512 or 1024 bits,

\item \texttt{FM-dummy2}, our variant(s) using the $(s,c,b,o)$ dense coding 
on nybbles, 
or $(c,b)$ dense coding on nybbles or triples of bits, with block sizes 
of 256 or 512 bits,

\item \texttt{FM-HWT2}, \texttt{FM-HWT4} and \texttt{FM-HWT8}, 
our variants based, respectively, 
on the 2-, 4- and the 8-ary Huffman-shaped wavelet tree, 
with block sizes of 512 or 1024 bits; 
note that for \texttt{FM-HWT2} part of the linear scan over 
a block is eliminated just like in the \texttt{FM-dummy1} and 
\texttt{FM-dummy2} variants denoted with letter `c' in their name.

\end{itemize}

As can be seen, our variants are significantly faster than existing 
implementations, yet they also take up much more space.
Among our variants the ones based on the multiary wavelet tree are most 
compact (except for the DNA case, where \texttt{FM-dummy3} is better 
in this aspect) and may often be preferred.
If more search speed is required, we can use \texttt{FM-dummy2cb}. 
We also note that \texttt{FM-dummy2}, using the more complex $(s,c,b,o)$ 
dense code, is never competitive.
Among \texttt{FM-dummy1}, \texttt{FM-dummy2} 
and \texttt{FM-HWT2} variants, the ones labeled 
with `c' were always faster (while using the same amount of memory) than 
their simpler counterparts, and we omitted showing the results from the 
slower implementations.

\begin{figure}
\centerline{
\includegraphics[width=0.49\textwidth,scale=1.0]{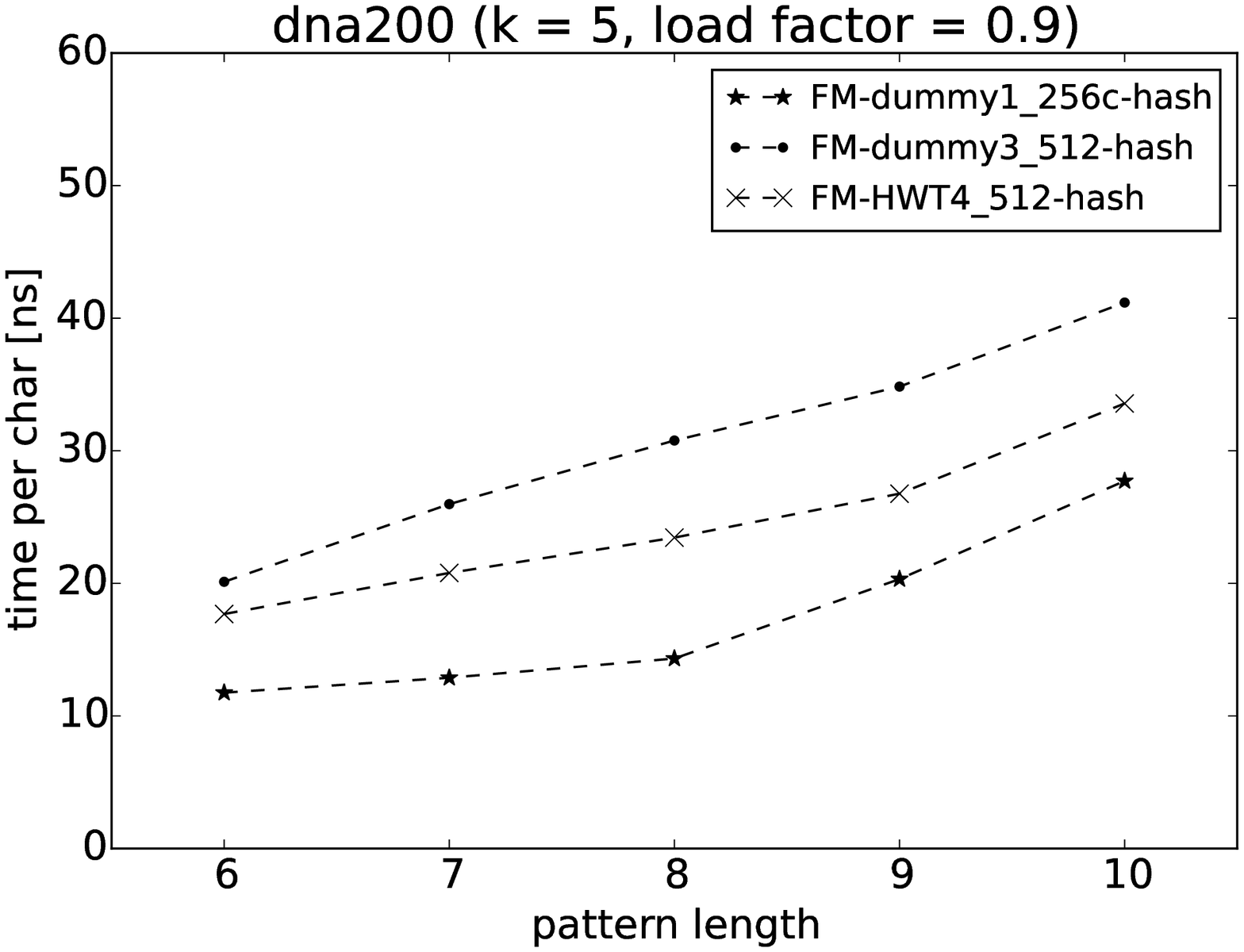}
\includegraphics[width=0.49\textwidth,scale=1.0]{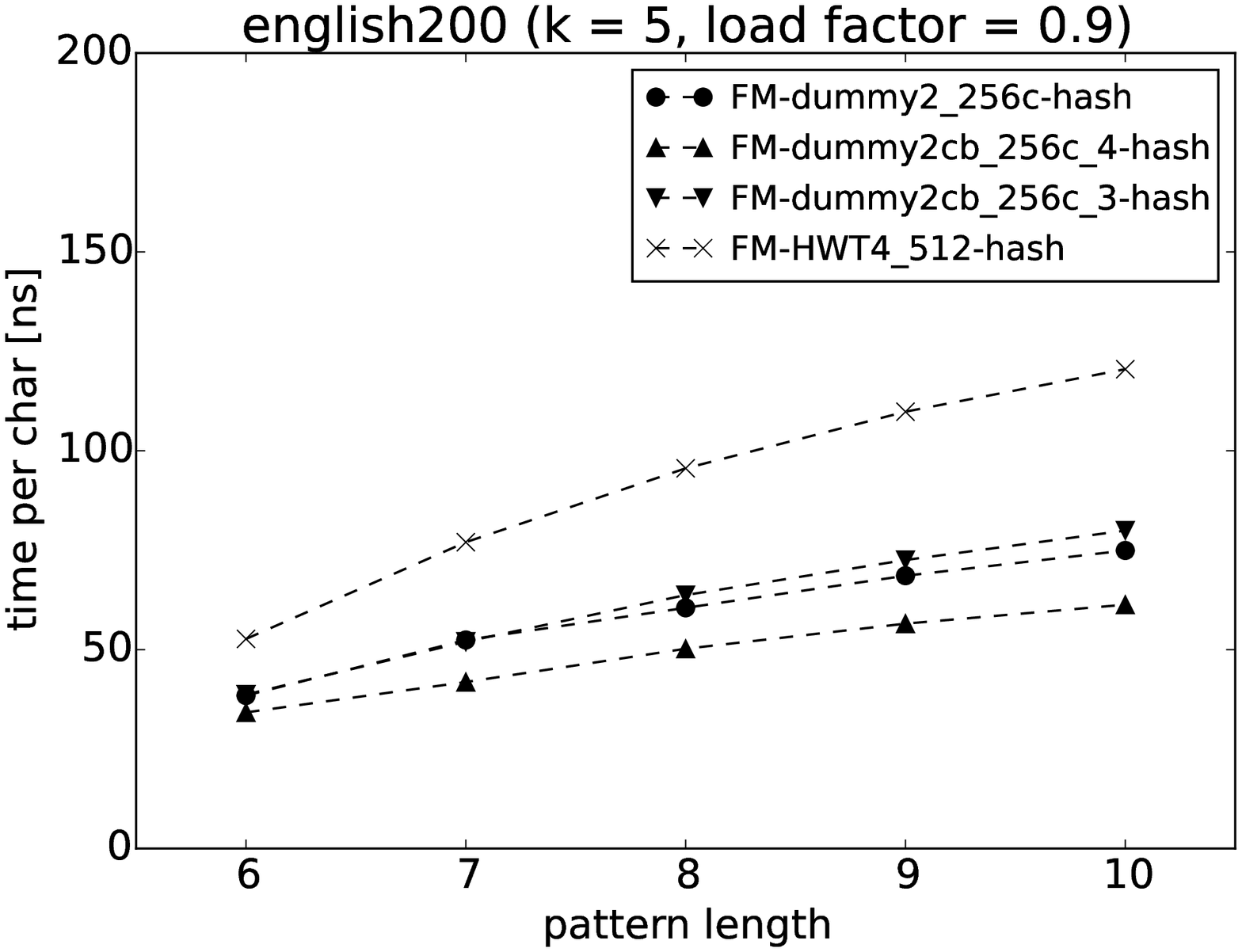}
}
\caption[Results]
{Count query for the variants with a hash table. 
1M patterns of length $\{6, 7, \ldots, 10\}$ were used. 
Times are averages in ns per character.
The patterns were extracted from the respective texts.
The extra space added by the hash table component was 
$0.0026n$ for \texttt{dna} and the variants 
\texttt{FM-dummy1} and \texttt{FM-dummy3},
$0.0029n$ for \texttt{dna} and \texttt{FM-HWT4},
and $0.1506n$ for \texttt{english} and all the variants.}
\label{fig:times-hash}
\end{figure}

In the next experiment we compared a few variants involving a hash table 
(see Section~\ref{sec:ht}).
The results, in Fig.~\ref{fig:times-hash}, are shown only for 
two datasets, but the trends are similar on the other Pizza~\&~Chili datasets 
as well.
The load factor for the hash tables was set to 0.9 in all cases.

\section{Conclusions}
We presented several simple FM-index variants, with preference 
to search speed rather than succinctness.
While most of the applied ideas are hardly novel, we believe some of them
have not been experimentally verified before.
Perhaps the most important building brick that we introduce is 
the (uncompressed) rank with one cache miss in the worst case.
Also, we note that Navarro in his survey on wavelet trees~\cite[p.~7]{Nav13} 
claims about the multiary variants of this data structure 
that ``although theoretically attractive, it is not easy to
translate their advantages to practice''.
Our results suggest however that Huffman-shaped 4- and 8-ary wavelet trees 
offer interesting space-time tradeoffs.

\section*{Acknowledgement}
We thank Simon Gog for helping us in using the sdsl-lite library 
and Shaun D.\ Jackman for an important remark concerning the ABySS 
de novo genome assembler.

The work was supported by the Polish National Science Centre upon decision 
DEC-2013/09/B/ST6/03117.

\bibliographystyle{abbrv}
\bibliography{fm}

\end{document}